\documentstyle[pra,aps,epsfig]{revtex}
\begin{document}

\draft

\title{Measurement of the $^{21}$Ne Zeeman frequency shift due to Rb-$^{21}$Ne
collisions}

\author{R.E. Stoner and R.L. Walsworth}
\address{Harvard-Smithsonian Center for Astrophysics, Cambridge, MA 02138}

\date{\today}

\maketitle

\begin{abstract}
We compared the frequency shift of the $^{21}$Ne Zeeman resonance
induced by polarized Rb vapor to the shift induced in the $^{3}$He Zeeman
resonance. The $^{3}$He/Rb shift has recently been measured with high
precision \cite{romalis98}, permitting the conversion of our differential measurement to
an absolute value for the $^{21}$Ne/Rb shift. We report a value of $\kappa _{21}=31.8\pm 2.8$ for
the Rb-$^{21}$Ne enhancement factor at a temperature of $124.8^o \pm 1.9$$^{o}$C. We propose 
high-precision differential contact shift measurements, the absolute accuracy of which would be
limited by the error in the $^{3}$He contact shift.
\end{abstract}



\section{Introduction}
\label{intro}
Polarization of noble gases by spin exchange with optically pumped
alkali metal vapors \cite{bouchiat60,walker97} is an important technique for a variety of
scientific and technological applications. Polarized noble gases been used for: precise
measurements of  the neutron spin structure function \cite{abe97}; neutron polarizers and
analyzers \cite{greene95}; magnetic resonance imaging
\cite{ng_imaging_i,ng_imaging_ii,ng_imaging_iii}; studies of gas diffusion in porous media
\cite{mair98}; gain media for Zeeman masers \cite{chupp94,stoner96,romalis99}; and tests of
fundamental symmetries \cite{bear98,chupp90,chupp89,vold84}.

New tests of fundamental symmetries using polarized noble gases have
recently been proposed. Kostelecky and Lane recently reported that a
$^{21}$Ne/$^{3}$He differential magnetometer could be used to test local Lorentz
invariance (LLI) \cite{kostelecky99}. We have proposed to develop a $^{21}$Ne/$^{3}$He dual
noble gas maser to carry out such an LLI test \cite{stoner99}.

The spin exchange interaction between alkali and noble gas atoms has
been extensively described in the literature
\cite{walker97,romalis98,walker89,happer87,happer72}. During spin exchange collisions, there is
significant overlap of the alkali atom's valence electron wavefunction with the noble gas
nucleus which comprises a Fermi contact interaction of the form
\begin{equation}
H_{int}=\alpha \vec K\cdot \vec S
\label{hyperfine_hamiltonian}
\end{equation}
where $\vec S$ and $\vec K$ are the electron and nuclear spin angular momentum operators,
respectively. This interaction leads to exchange of angular momentum between the alkali valence
electron and the noble gas nucleus. The valence electron and nucleus experience during the
collision a strong magnetic field from the close proximity of their spin magnetic moments. In
spite of the small fraction of the time spent in collision, the average magnetic fields
experienced by both the alkali valence electron and the noble gas nucleus
can be affected if either the alkali or noble gas species is polarized.
The resultant change in the Larmor precession frequency for either species, referred to as the
contact shift, is expressed in terms of an
enhancement factor $\kappa$.
$\kappa$ is defined in terms of the \emph{total} shift in the Larmor precession
frequencies in co-located spherical distributions of magnetizations, induced both by the contact
interaction and by the classical magnetic fields generated by the magnetizations. $\kappa$ is the
ratio of that total frequency shift (contact plus classical magnetization) to the shift that would
be induced by the classical magnetization alone.
$\kappa$ can be much greater than unity, increasing with increasing $Z$ of the noble gas nucleus
\cite{walker89}. In the limit where the formation of alkali-noble gas van der Waals molecules is a
negligible part of the spin exchange interaction (true for the experiment reported in this paper),
a single enhancement factor describes both the contact shift of the noble gas Zeeman splitting by
the alkali atoms, and vice versa.
\cite{schaefer89}. The frequency shift experienced by noble gas atoms in the presence of polarized
alkali atoms, including both the contact shift and the shift due to classical magnetization
fields, is
\begin{equation}
\delta \omega _{ng}=\left( {\kappa -1+f_{DDF}} \right)\gamma _{ng}\cdot {{8\pi } \over 3} g_S \mu
_B\left[ a \right]\left\langle {S_z} \right\rangle 
\label{delta_omega_vs_m}
\end{equation}
Recall the elementary result that ${{8\pi } \over 3}M$ is the magnetic field strength inside a
magnetized sphere with magnetic dipole moment density $M$; the field is uniform and parallel to
the magnetization. $f_{DDF}$ is a dimensionless factor that accounts for the
effect of classical magnetization fields ("distant dipole fields" \cite{romalis98}), and is unity
for a spherical distribution of magnetization. $\left[ a \right]$ is the number per unit volume of
alkali atoms, and $\gamma_{ng}$ is the gyromagnetic ratio of the noble gas species. For
convenience, we define the component of alkali atom magnetization along the magnetic field (z)
axis, 
$M_{a}$, as
\begin{equation}
M_{a}=\hbar \gamma _{a}\left[ {a} \right]\left\langle {S_z} \right\rangle 
\label{m_vs_p}
\end{equation}
where $\gamma _a={{g_S \mu _B} \mathord{\left/ {\vphantom {{g_S\mu _B} \hbar }} \right.
\kern-\nulldelimiterspace} \hbar } =$ $2 \pi \cdot$1.40 MHz/gauss is the electron gyromagnetic
ratio. The alkali nuclear magnetic moments do not contribute to the contact shift, but do affect
the distant dipole fields produced by the alkali magnetization distribution. We ignored these
effects on the distant dipole fields, since they were too small to bear on the results reported in
this work. 
 
In precision measurements, it is important to be able to monitor the
absolute polarization of the noble gases. Romalis and Cates \cite{romalis98}
demonstrated accurate polarimetry of $^{3}$He
using the Rb Zeeman frequency
shift due to Rb-$^{3}$He spin exchange collisions, and made a very precise
measurement of the Rb-$^{3}$He enhancement factor. Future use of $^{21}$Ne in precision
measurement  applications would benefit from the use of contact shift polarimetry, but there is
at present no published measurement of the Rb-$^{21}$Ne enhancement factor. In this paper we
present the first such measurement. We measured
the net Larmor frequency shifts, simultaneously impressed by the presence of polarized Rb, on
ensembles of polarized
$^{3}$He and $^{21}$Ne precessing in the same [nearly] spherical glass cell. The measurement
presented in this paper is not the first in which a ratio of contact shift enhancement factors was
determined. Baranga \emph{et. al.} measured the shifts of the Larmor precession
frequencies of Rb and K vapors residing in the same cell, induced simultaneously by the presence
of polarized ${}^3$He \cite{baranga98}.

The experiment is described in Section II. Determining the ratio of the
Rb-induced shifts permitted us to divide out the dependence on Rb magnetization, and express the
${}^{21}$Ne enhancement factor 
$\kappa_{21}$ in terms of the ${}^3$He enhancement factor $\kappa_{3}$. The very accurate
measurement of
$\kappa_{3}$ by Romalis and Cates determined the absolute contact shift enhancement factor
$\kappa_{21}$. Analysis of the data is presented in Section III. Suggestions for improved
measurements are given in Section IV.

\section{Apparatus, procedure, system parameters}
\label{apparatus_section}
\begin{figure}[]
\begin{center}
\epsfig{file=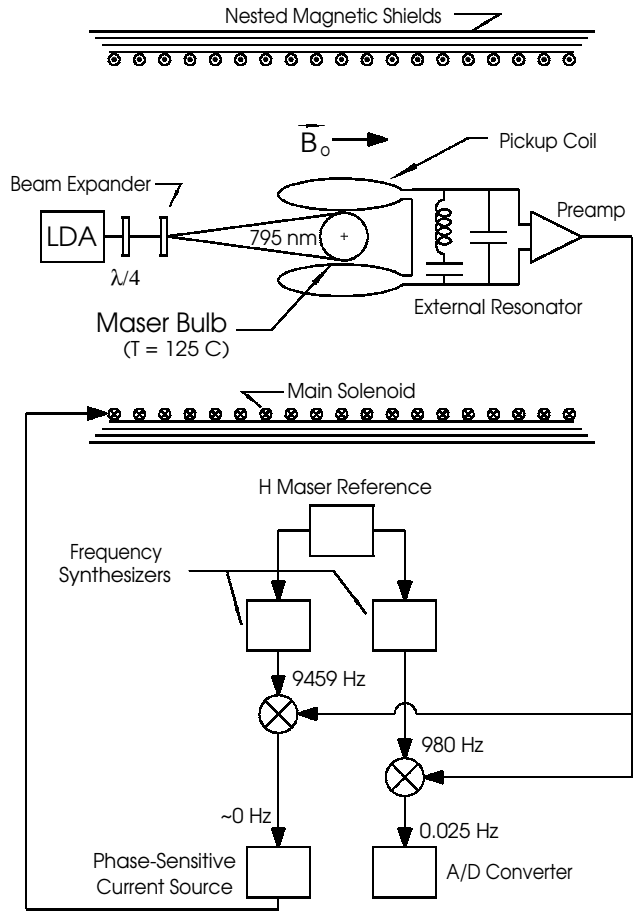,height=5.5in,width=4.25in, bbllx=0in,bblly=0in,bburx=8.5in,bbury=11in}
\end{center}
\caption{The 
${}^{21}$Ne/${}^{3}$He dual Zeeman maser. The device operates similarly to the
${}^{129}$Xe/${}^{3}$He maser \protect\cite{stoner96}, except that the ${}^{21}$Ne/${}^{3}$He
maser employs a single- rather than a double-bulb cell. Co-located ensembles of noble gas atoms
are polarized by spin exchange collisions with optically pumped Rb atoms. Precession of
polarized atoms excites current in nearby inductive pickup coils, which are part of a circuit
tuned with resonances near the Larmor precession frequencies of the two noble gas species. The
current induces magnetic fields which act back on the precessing atoms, providing feedback for
maser oscillation.}
\label{fig1}
\end{figure}

We employed a $^{21}$Ne/$^{3}$He dual noble gas Zeeman maser to establish
simultaneous precession of the $^{21}$Ne and the $^{3}$He in the same sealed
glass cell. The $^{21}$Ne/$^{3}$He maser is a new device that will be described
elsewhere; that description is summarized here. Figure \ref{fig1} is a
schematic representation of the maser. Circularly polarized light resonant
with the Rb D1 transition (795 nm) is generated by a 30 W laser
diode array (LDA); the light is directed onto a 1.4 cm dia [near-] spherical
sealed glass cell containing 228 torr of $^{21}$Ne, 266 torr of $^{3}$He, 31 torr of N${}_2$,
and Rb metal of natural isotopic composition \cite{chupp_pc}. The LDA light propagates along
the axis of a uniform magnetic field of strength 3 Gauss. The field is generated by a precision
solenoid/$\mu$-metal-shield system capable of producing a 3 G magnetic field with $\le 35$
$\mu$G/cm magnetic field gradients, with no gradient shimming. The main solenoid was driven
using a current source capable of $\sim 3$ ppm stability over timescales of order 1000 sec. The
glass cell was heated both by the LDA light and blown hot air to a temperature of
($124.8 \pm 1.9$)${}^o$C (see Sec. III). The
temperature of the air surrounding the cell was controlled to about 43 mK RMS. The total output
power of the LDA was actively controlled to about a part in $10^4$. 

The Rb was polarized by optical pumping; the resulting Rb electron polarization was very close to
unity.  The output power of the LDA was about
30W, so that even though it had a very broad ($\sim 1.5$ nm) linewidth, the on-resonant optical
power was roughly 0.5 W. Thus, the optical pumping rate was $>10^5$ photon absorptions per sec per
Rb atom. The rate of Rb polarization destruction events per Rb atom in the cell was $<10^3$
sec${}^{-1}$ so that the Rb electron polarization was greater than 0.99. Pickup coils wound in
series were placed in close proximity to the cell. The coils were excited by the magnetic flux
associated with the precession of the polarized noble gases. The coils were part of a tuned
resonant circuit with resonances at both noble gas precession frequencies. The large current flow
induced in the pickup coils by resonant excitation resulted in magnetic fields oscillating at the
noble gas nuclear Larmor frequencies being impressed back on the precessing atoms. This feedback
resulted in steady state maser operation.

$^{21}$Ne has nuclear spin 3/2, and thus a quadrupole moment. The quadrupole moments of the
$^{21}$Ne  ensemble interact coherently with electric field gradients at the glass cell walls,
shifting the four Zeeman energy levels so as to split the three [otherwise degenerate] dipole
transitions \cite{zu88}. The splitting is proportional to
$\delta \omega _Q\propto 3\cos ^2\theta -1$, where $\theta$ is the angle of the cell's symmetry
axis with the magnetic field. We oriented the cell as close as possible to the "magic angle"
$\cos \theta =\sqrt {{1
\mathord{\left/ {\vphantom {1 3}} \right. \kern-\nulldelimiterspace} 3}}$
to minimize the quadrupole splitting. The quadrupole wall shifts in the cell were not
measured; however, computer simulations indicate that without proper cell orientation,
realistic-sized splittings could disrupt $^{21}$Ne maser oscillation. In any case,
$^{21}$Ne maser oscillation with steady-state amplitude was attained. We observed a small ($\sim
10$ $\mu$Hz) sinusoidal modulation of the $^{21}$Ne maser frequency. The period of the
modulation was about 16 hr. We speculate that the modulation was a consequence of remnant
quadrupole splitting resulting from cell orientation error. In any case, the effect was small,
with a period much longer than the duration of our contact shift measurements, and thus was not a
source of significant systematic error.

The $^{3}$He and $^{21}$Ne maser signals were presented to a low-noise preamplier. The
preamp output was then analyzed using lockin amplifiers, which were
referenced to a set of synchronized signal generators tuned near the noble gas maser frequencies
and locked to an H-maser-derived 5 MHz signal. The phases of both noble gas masers' precession were
recorded digitally at a sample rate of 1 Hz. The A/D sampling trigger was
also derived from the master 5 MHz reference.  

The measurements consisted
of observing the maser phases during a series of reversals of the
direction of the Rb polarization. The Rb polarization was reversed by rotation of  
the quarter-wave plates (Fig. 1) and was typically achieved in $\le $
20 sec. The total duration of
a set of reversals was about 1000 sec., limited by magnetic field stability as well as the
extent to which the noble gas masers' steady states were disrupted by the alternating Rb
polarization. Two sets of reversal data were acquired. Clear changes in the slopes of the phase
curves were precisely correlated with reversal of the Rb polarization direction; these frequency
changes were the result of the contact shift combined with the [classical] shift due to the Rb
magnetization distribution. The change of noble gas maser oscillation frequency induced by Rb
polarization reversal was about 44 mHz for $^{21}$Ne maser and 71 mHz for the $^{3}$He maser.

\section{Data and analysis}
\label{data_analysis_section}
\begin{figure}[]
\begin{center}
\includegraphics{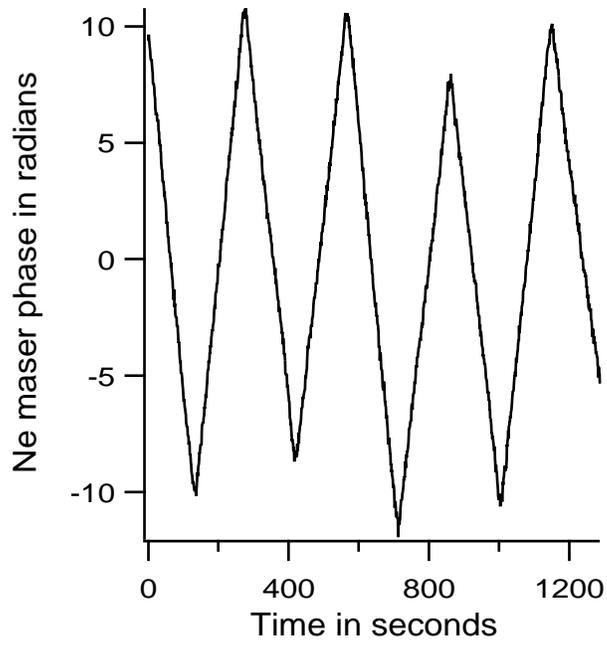}
\end{center}
\caption{Example ${}^{21}$Ne maser phase profile from one (of two) measurement scans. Phase
dependence linear in time has been removed.}
\label{fig2}
\end{figure}

\begin{figure}[]
\begin{center}
\includegraphics{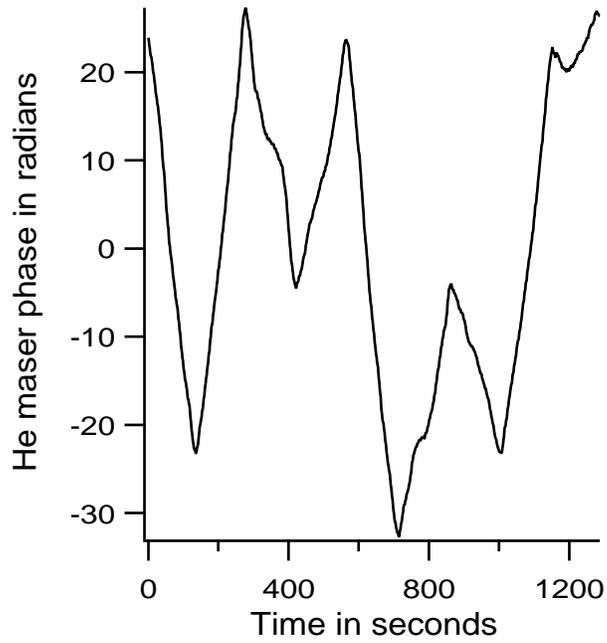}
\end{center}
\caption{Example ${}^{3}$He maser phase profile measured simultaneously with the data of Fig. 2.
Phase dependences linear in time have been removed. Note the considerable distortion in this
phase curve due to magnetic field drift.}
\label{fig3}
\end{figure}

Phase data for the second of the two measurement scans are shown in Figs. \ref{fig2} and
\ref{fig3}. We model the
${}^{21}$Ne phase $\varphi _{21}$ (see Fig. \ref{fig2}) and the ${}^{3}$He phase $\varphi _{3}$
(see Fig. \ref{fig3}) acquired by the lockin amplifiers as evolving according to the following
equations:
$$\varphi _{21}=\varphi _{21,o}+\gamma _{21}B_ot-2\pi \upsilon _{ref,21}+\gamma _{21}\int_0^t 
{dt'\delta B\left( {t'} \right)}+\left( {\kappa _{21}-1+f_{DDF}} \right)\gamma _{21}{{8\pi }
\over 3}\int_0^t {dt'M_{Rb}\left( {t'} \right)}$$
\begin{equation}
\varphi _3=\varphi _{3,o}+\gamma _3B_ot-2\pi \upsilon _{ref,3}+\gamma _3\int_0^t {dt'\delta 
B\left( {t'} \right)}+\left( {\kappa _{3}-1+f_{DDF}} \right)\gamma _3{{8\pi } \over
3}\int_0^t {dt'M_{Rb}\left( {t'} \right)}
\label{phi_defs}
\end{equation}
where: $\gamma_{21}$ ($\gamma_3$) is the gyromagnetic ratio of Ne (He); $\kappa_{21}$
($\kappa_3$) is the enhancement factor for Ne (He); $f_{DDF}$ is the dimensionless
factor to account for the classical magnetic field due to the Rb magnetization;
$\nu_{ref,21}$ and $\nu_{ref,3}$ are the reference frequencies presented to the lockin amplifiers;
$B_o$ is the time-average magnetic field strength over the time interval; $\delta B$ is the
time-dependent deviation of the magnetic field from its mean value; and $M_{Rb}$ is the Rb
magnetization. Note that independent maser frequency measurements determined the ratio of the
two gyromagnetic ratios, ${{\gamma _3} \mathord{\left/ {\vphantom {{\gamma _3} {\gamma _{21}}}}
\right. \kern-\nulldelimiterspace} {\gamma _{21}}}$, to an accuracy much better than 1 ppm.
Also, the effects of magnetic fields due to noble gas magnetization fields on the phases
have been neglected in comparison to the contact shift effects; this is reasonable for the
noble gas magnetizations of this experiment (see below).

\begin{figure}[]
\begin{center}
\includegraphics{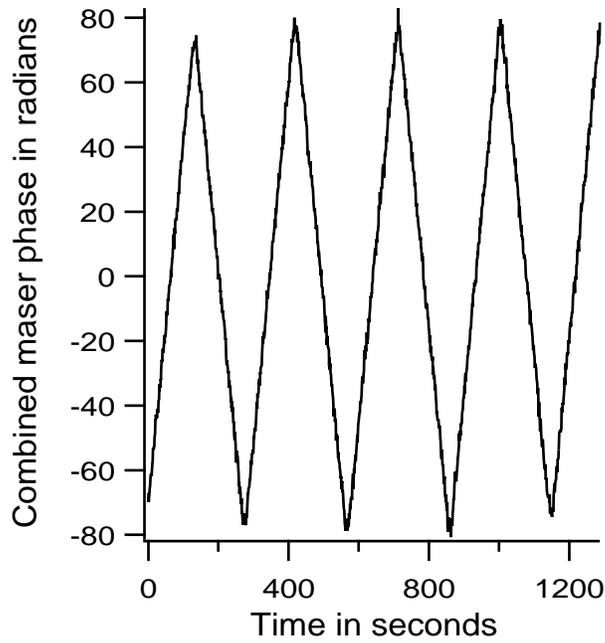}
\end{center}
\caption{The combined phase $\Delta \varphi$ of the two masers as per
eqn (\protect\ref{delphi_def}). This profile is proportional to the phase variation induced by the
Rb-noble gas contact shift alone. The sawtooth shape arises from periodic reversal of the Rb
polarization, which reverses the Rb-noble gas contact frequency shift. The highly regular shape of
the difference curve reflects the fact that magnetic field drift effects are subtracted out.}
\label{fig4}
\end{figure}

Magnetic field drift occurred over the course of the measurement scan. However, the ${}^{21}$Ne and
${}^{3}$He masers are co-located magnetometers, so that we can construct a data set from which
the effects of magnetic field drift and Rb distant dipole fields have been removed. We
define
\begin{equation}
\Delta \varphi \equiv \varphi _3-{{\gamma _3} \over {\gamma _{21}}}\varphi _{21}
\label{delphi_def}
\end{equation}
It is easily shown that
\begin{equation}
\Delta \varphi =\varphi _{3,o}-\varphi _{21,o}+2\pi \left( {{{\gamma _3} \over
{\gamma _{21}}}\nu _{ref,21}-\nu _{ref,3}} \right)t+\left( {\kappa _3-\kappa _{21}}
\right)\gamma _3{{8\pi } \over 3}\int_0^t {dt'M_{Rb}\left( {t'} \right)}
\label{delphi}
\end{equation}
Aside from a trivial phase offset and known linear phase evolution, $\Delta \varphi$ reflects
the action of the Rb contact shifts only: phase evolution due to magnetic fields is subtracted off.
This combined phase profile is plotted in Fig. \ref{fig4}. As seen in Fig.
\ref{fig4}, the phase evolution due to the contact shift was a piecewise-linear "sawtooth"
profile. Since our measurement is derived from computing the differences in slopes between
adjacent linear regions of the curve, removal of constant and linear dependence from the phase
data had no effect on the measurement. Thus, our plots of the phase data shown in Figs.
\ref{fig2}, \ref{fig3}, and \ref{fig4} are displayed with mean value and mean slope of zero.

If the Rb magnetization were known, one could immediately deduce $\kappa_{21}$ using eqn
(\ref{delphi}). Even though the Rb electron spin polarization was very close to unity,
the density was not known \emph{a priori}. It was therefore necessary to consider the
ratio of contact shifts impressed on the $\varphi _{21}$ and the $\Delta \varphi$ phase curves,
and then determine the temperature self-consistently.

The data were analyzed in two ways which yielded essentially identical results. First, we 
computed the average changes in slope of the phase curves $\varphi _{21}$ and $\Delta \varphi$,
caused by the reversals of the Rb magnetization. That is, a least-squares linear fitting
routine was used to extract the slope of each piecewise linear region of the phase curves. The
differences in slope of adjacent regions was then recorded for each of the phase curves. 
The mean absolute values of the slope differences were then computed for each data set.
We define $M_{21}$ as the mean of the absolute value of the slope differences for
$\varphi_{21}$, and $\Delta M$ as that for the $\Delta \varphi$ profile. The uncertainties were
estimated as the standard deviation of the mean in the slope differences' absolute values.

We define the ratio of mean slope changes as ${\mathcal R}\equiv{{M_{21}} \mathord{\left/
{\vphantom {{M_{21}}  {\Delta M}}} \right. \kern-\nulldelimiterspace} {\Delta M}}$. The
weighted average value for ${\mathcal R}$ from the two data sets was ${\mathcal
R}=0.1262 \pm 0.0025$ where a one-sigma uncertainty is reported.  The uncertainty was almost
entirely due to that in $M_{21}$.  The fractional uncertainty in $\Delta M$ for both of the data
sets was $< 0.1\%$ This indicates that Rb polarization reversal was achieved to high precision
and that the Rb magnetization was stable to better than 0.1$\%$ over the course of the
measurements. The fractional uncertainty in $M_{21}$ was much larger, $\sim 2\%$, caused by the
effects of magnetic field drift. In principle, we could have analyzed the
$\varphi_{3}$ profile rather than $\varphi_{21}$, but magnetic field drift effects were
${{\gamma _3}
\mathord{\left/ {\vphantom {{\gamma _3} {\gamma _{21}=9.65}}} \right.
\kern-\nulldelimiterspace} {\gamma _{21}=9.65}}$ times larger on $\varphi_3$ than
$\varphi_{21}$, making it impractical to extract slope data from $\varphi_3$ directly (see Fig.
\ref{fig3}).

In our analysis we did not account for possible variations in frequency due to changes in the 
noble gas longitudinal polarization that occurred in the course of a measurement. The
magnetization of the noble gases created magnetic fields which varied in time during
the course of the each of the two measurement scans, as evidenced by [small] maser amplitude
changes  observed during the scans. A given maser's magnetization field could not affect its own
frequency, since magnetization fields were parallel to the magnetization in the near-spherical
cell. However, each maser's frequency was shifted by the other's magnetization, and such shifts
did not subtract out of $\Delta \varphi$ as did magnetic field drift. The noble gas polarizations
underwent both slow drift as well as variations correlated with the Rb polarization reversals. The
small uncertainty in the $\Delta \varphi$ slope differences proves that frequency variations due to
slow drift in the noble gas magnetizations had negligible effect. We also determined that the
variations of the noble gas polarization caused by the periodic Rb polarization reversal
had negligible effect on the contact shift measurement. Periodic reversal
of the Rb polarization  (i.e., "square wave" modulation of the Rb polarization) induced a "sawtooth
wave" modulation (i.e., integrated square wave) on the noble gas polarizations of peak-to-peak
amplitude approximately $\gamma_{SE} \tau$, where $\gamma _{SE}$ is the Rb-noble gas spin exchange
rate per noble gas atom, and $\tau$ = time between Rb polarization reversals. 
Thus, the varying noble gas
polarizations induced large changes in the maser frequencies over a measurement period: we
estimate that a 2.8 mHz shift was induced on the ${}^3$He maser by the ${}^{21}$Ne polarization
variation and a 0.7 mHz shift was induced on the ${}^{21}$Ne maser by the ${}^3$He polarization
variation. However, given a constant time interval $\tau$ between Rb polarization reversals, the
\emph{average} frequency shifts due to the Rb-driven noble gas polarization variations did not
change from one period to the next, thus they subtracted out in the calculation of the
period-to-period frequency changes $M_{21}$ and $\Delta M$. We conclude that noble gas
polarization-induced magnetic fields had no systematic effect on the measurement of ${\mathcal R}$.

It is easy to show from eqns (\ref{phi_defs}) and (\ref{delphi}) that the ratio 
${\mathcal R}$ can be interpreted as
\begin{equation}
{\mathcal R}=\left| {{{\gamma _{21}} \over {\gamma _3}}\cdot {{\kappa _{21}+\left( {f_{DDF}-1}
\right)}
\over {\kappa _3-\kappa _{21}}}} \right|
\label{r_defn}
\end{equation}
where it is seen that dependence on the Rb magnetization divided out. The value and
estimated uncertainty for ${\mathcal R}$ can then be used to compute the value and uncertainty
of the desired contact shift enhancement factor for ${}^{21}$Ne,
$\kappa _{21}$, though the system temperature must be known because ${}^3$He contact shift
$\kappa _{3}$ is temperature-dependent. $\kappa _{21}$ and the system
temperature must be determined self-consistently, and the effect of classical magnetization
fields must be accounted for: these questions are addressed below.

We now consider the second method of analyzing the data to obtain a value for ${\mathcal R}$, which
yielded essentially the same results as the above approach. While this second method is
slightly more complex, it is useful because it provides a framework for estimating the size
of magnetic field drift in the present measurement. This second method of extracting a value of
${\mathcal R}$ from the data considers the phase data sets to be $N$-component vectors (denoted by
a tilde), where
$N \equiv$ number of data in the set ($\sim 10^3$ for each of the two scans). We use the
$\Delta \varphi$ profile as a model of the shape of the Rb
magnetization-induced phase (see eqn \ref{delphi}). We then seek to describe the $\tilde
\varphi _{21}$ profile vector as a linear combination of elements of a partial basis of
orthonormal vectors. The partial basis included an element corresponding to the [normalized]
$\Delta \varphi$ profile ($\equiv \tilde V_o$). Additional basis vectors orthogonal to it,
$\left\{ {\tilde V_i} \right\}_{i=1,N_b-1}$ were derived from $\Delta \varphi$ and Legendre
polynomials $\left\{ {P_i} \right\}_{i=0,{N_b-1}}$ using Gram-Schmidt orthogonalization
\cite{hoffman_kunze}. The number of basis vectors $N_b$ was of order $10^1$, whereas the
dimension of the vector space $N$ was of order $10^3$, so $N_b<<N$, and we find:
\begin{equation}
\tilde \varphi _{21}=\sum\limits_{i=0}^{N_b-1} {a_i\tilde V_i}
\label{lin_comb}
\end{equation}
The coefficient of $\tilde V_0$ in this linear combination, $a_0$, was
proportional to the value for ${\mathcal R}$ obtained from that data set. The coefficients of
the linear combination were determined using linear least squares fitting and by simple
projection of $\tilde \varphi _{21}$ onto the vectors of the partial basis, with identical
results. The number of elements in the basis set was varied, but the coefficients obtained from
the least squares fit remained the same regardless of the size of the basis set.

Using this second data analysis method, we thus determined two values of ${\mathcal R}$, one from
each data set. It was not possible to estimate uncertainties for the individual values of
${\mathcal R}$ from the vector analysis method. Uncertainty estimates from the least squares
analyses were unrealistically small because the vector analysis could not account for the effect
of a non-zero projection of the magnetic field drifts along the Rb magnetization-induced phase
profile vectors. We take the value of ${\mathcal R}$ to be the mean of the results for the two
data sets, and we coarsely estimate the uncertainty of ${\mathcal R}$ as the deviation of the two
values, to obtain
${\mathcal R}=0.1287 \pm 0.0036$. This result agrees well with that of the first method of data
analysis. For computing
$\kappa _{21}$ we will use the result from the first method of analysis since the error estimate
is probably more reliable.

\begin{figure}[]
\begin{center}
\includegraphics{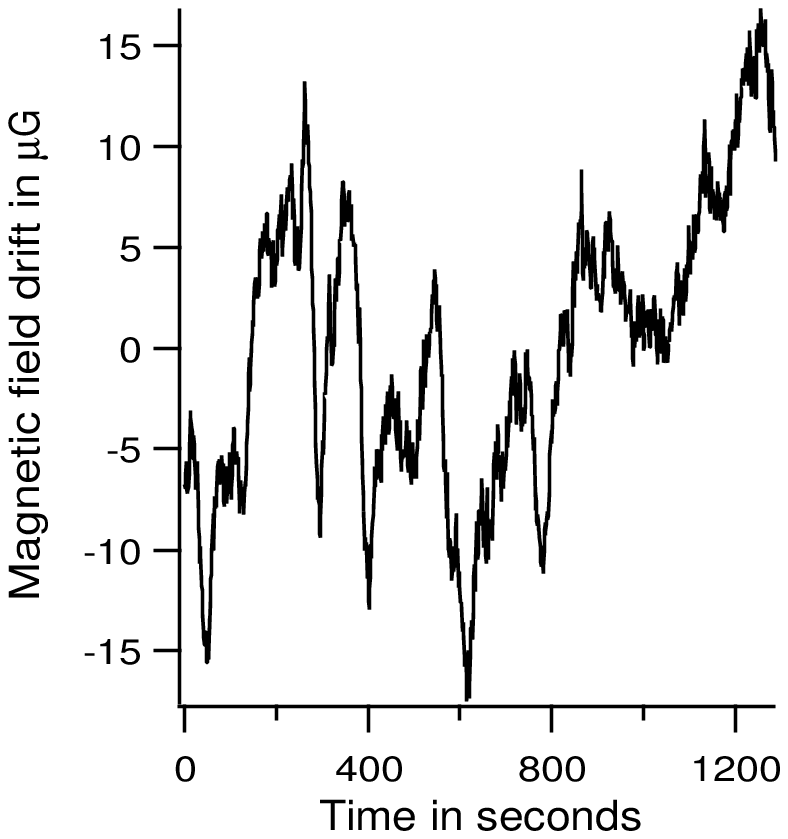}
\end{center}
\caption{Magnetic field drift during the measurements of Figs.
\protect\ref{fig2} and \protect\ref{fig3}, computed from eqn (\protect\ref{b_drift}). The field
strength was 3 G, so that the fractional RMS stability of the magnetic field was about 3 ppm.
Field drift was the dominant source of experimental error.}
\label{fig5}
\end{figure}

We can compute the magnetic field drift profile by subtracting from $\tilde \varphi_{21}$ its
projection on the Rb magnetization basis profile $\tilde V_0$:
\begin{equation}
\tilde {\delta B}={1 \over {\gamma _{21}}}{d \over {dt}}\left( {\tilde \varphi
_{21}-\left( {\tilde \varphi _{21}\cdot \tilde V_o} 
\right)\tilde V_o} \right)
\label{b_drift}
\end{equation}
Note that the profile $\tilde {\delta B}$ is the magnetic field drift
\emph{orthogonal} to $\tilde V_0$. The drift profile from the second data set is plotted in Fig.
\ref{fig5}; the time derivative of the [discretely sampled] profile was estimated by finite
difference. Note that the field was stable to about
$\pm 3$ ppm RMS. 

Having obtained a value for ${\mathcal R}$, we can then solve eqn (\ref{r_defn}) for
$\kappa_{21}$ in terms of measured parameters, since $\kappa_3$ is known \cite{romalis98}. However,
$\kappa_{21}$ is expected to be temperature-dependent \cite{romalis98,walker89} and so the
temperature at which the experiment was carried out must be reported along with the value of
$\kappa _{21}$. The Rb and noble gas temperature indicated by the hot air temperature control
system was in error, because: the absolute calibration was poorly known; and the cell was heated
by the LDA to a temperature higher than that indicated by the control system. Thus, it was
necessary to deduce the noble gas/Rb vapor temperature from the contact shift data, and the known
${}^3$He enhancement factor (with its known temperature dependence), in a self-consistent way.
This was achieved by first noting that the mean slope change $\Delta M$ is related to the
enhancement factors and the Rb magnetization via
\begin{equation}
\Delta M=2\gamma _3{{8\pi } \over 3}\left| {\left( {\kappa _3-\kappa _{21}} 
\right)M_{Rb}\left( T \right)} \right|
\label{delta_m_vs_m_rb}
\end{equation}
where $\left| {M_{Rb}\left( T \right)} \right|$ (see eqn (\ref{m_vs_p}) above) is a
known function of temperature. The Rb number density $\left[ {Rb} \right]$
is related to temperature by \cite{alcock84}
\begin{equation}
\left[ {Rb} \right]={{10^{9.318-{{4040} \over T}}} \over {\left( {1.38\cdot 10^{-17}} 
\right)\cdot T}}cm^{-3}\ \ \left( {T\ in\ {}^oK} \right)$$
\label{n_rb_vs_t}
\end{equation}
The measured temperature dependence of the ${}^3$He enhancement factor is \cite{romalis98}
\begin{equation}
\kappa _3=4.52+0.00934\left[ {T\left( {{}^oC} \right)} \right]
\label{kappa_3_vs_t}
\end{equation}
Substituting (\ref{n_rb_vs_t}) and (\ref{m_vs_p}) into (\ref{delta_m_vs_m_rb})
yields an equation relating $\kappa_{21}$ and temperature (and known/measured parameters).
Substituting (\ref{kappa_3_vs_t}) into (\ref{r_defn}) yields another equation relating
$\kappa_{21}$ and temperature. This system of two [nonlinear] equations and two unknowns can be
solved to yield the temperature and $\kappa_{21}$. The temperature determination depends
strongly on the enhancement factor, while the enhancement factor determination depends only
weakly on the temperature (via the known temperature dependence of $\kappa_3$). Thus, an
iterative method to compute the temperature rapidly converges: an initial guess for the
temperature enables calculation of
$\kappa_{21}$, which in turn permits calculation of the Rb magnetization and thus the system
temperature. This corrected temperature is used to re-compute a better estimate for
$\kappa_{21}$, which is used to re-compute a better estimate of the Rb magnetization, etc.

  Using the result ${\mathcal R}=0.1262 \pm 0.0025$ from the slope difference analysis, we
determine the contact shift enhancement factor for Rb and ${}^{21}$Ne to be
$\kappa_{21} = 31.8
\pm 2.8$ at a  temperature of $(124.8 \pm 1.9)^o$C. Note that the noble gas/Rb vapor temperature is
much higher than the value of $112^o$C indicated by the hot air temperature control system,
presumably the result of heating by the laser diode array. We are not aware of a previous
measurement of the Rb-${}^{21}$Ne contact shift. Walker estimated the Rb-${}^{21}$Ne enhancement
factor to be 38, and the ratio of the Rb-${}^{21}$Ne and Rb-${}^{3}$He enhancement factors to be
4.3
\cite{walker89}. We report this ratio to be $5.6 \pm 0.5$.

  The $3\%$ uncertainty in ${\mathcal R}$ dominates other sources of error, inducing a $9\%$
uncertainty in  $\kappa_{21}$. The contribution to the uncertainty in $\kappa_{21}$ resulting
from the $1.5\%$ uncertainty in the He enhancement factor $\kappa_3$ is only $1.5\%$, which is
negligible. The uncertainty in temperature induced a  $0.3\%$ uncertainty in $\kappa_{21}$. To
the extent that the cell could be described as an ellipsoid of revolution, its deviation from
sphericality induced an uncertainty of $0.15\%$ in $\kappa_{21}$ with zero systematic correction. 

  The claim of zero systematic correction of the classical magnetization field shift
for a ellipsoidal cell is justified  by noting that: i) the classical magnetization field
inside an ellipsoid of uniform magnetization is uniform, with its magnitude depending on the
ellipsoid's shape and the relative orientation of the ellipsoid major axis and the
magnetization vector; and ii) when the ellipsoid major axis is oriented at the magic angle 
$\theta _m=\cos^{-1}\left( {{1 \mathord{\left/ {\vphantom {1 {\sqrt 3}}} \right.
\kern-\nulldelimiterspace} {\sqrt 3}}} \right)$ with respect to the magnetization direction, the
component of the classical magnetization field parallel to the magnetization is exactly that of
a spherical distribution with the same magnetization, regardless of the ellipsoid's shape. We
oriented the cell at
$\theta _m$ in order to suppress quadrupole wall shifts, with an uncertainty of about $10^o$.
The $10^o$ uncertainty in orientation leads to the estimated uncertainty of $0.15\%$ in
$\kappa_{21}$ due to measured asphericality of the cell.
 
\section{Suggestions for improved measurements; conclusions}
\label{conclusions_section}
There are straightforward improvements that could be implemented to reduce the uncertainty in 
the present measurement. The experiment was conducted in the course of developing a
${}^{21}$Ne/${}^{3}$He dual Zeeman maser, so that opportunity for taking contact shift data
was very limited. A factor of three improvement in the result could easily be achieved just by
taking ten times more data. Active field stabilization using an independent magnetometer
\cite{romalis98} would essentially eliminate field drift, reducing statistical uncertainty in
the measurement to that due to phase noise, a factor of $> 0.03/0.001 = 30$ reduction. Active
field stabilization would also permit a reduction in the phase noise since the signal
acquisition bandwidth could be reduced by at least a factor of 3, so that the statistical
uncertainty could be reduced by about a factor of 300 compared to the present experiment. This
uncertainty would already be less than the uncertainty contributed by cell
asphericality ($0.15\%$). However, the absolute uncertainty in
the Rb-${}^{21}$Ne contact shift enhancement factor
$\kappa_{21}$ would then be limited by the uncertainty in that for He, $\kappa_{3}$, $1.5\%$,
whereas the ratio $\kappa_{21}/\kappa_{3}$ would be known to an uncertainty of only a few tenths
of one percent.

It should be possible to measure ratios of alkali-noble gas
enhancement factors to a level well below a part in $10^3$ if one could completely eliminate
effects of the classical magnetization field shift. We have already shown that one can use
differential magnetometry to obtain data from which classical magnetization field shifts are
eliminated to high precision. It can be shown that using a three-species noble gas maser, one
could measure the contact shift enhancement factor of a third species if the other species'
enhancement factors are known, \emph{independent of the classical magnetization field shift}. Even
though it has not yet been done, we believe it is a straightforward matter to implement a
three-species maser in a single-bulb glass cell.

A comprehensive set of contact shift data would serve as a probe of noble gas-alkali
interaction potentials \cite{walker89}. A program using multi-species noble gas masers to measure 
contact shift ratios could be
implemented using ${}^{129}$Xe, ${}^{83}$Kr, ${}^{21}$Ne, and ${}^{3}$He along with K, Cs, and
Rb in turn. Na would be a poor candidate for use in a noble gas maser because of its relatively
low vapor pressure. The remaining stable noble gas isotope, ${}^{131}$Xe, cannot be used with the
other noble gases in a multi-species maser because its nuclear dipole moment has a sign opposite to
that of the others (making it impossible to achieve a population inversion simultaneously with
the other species). However, it may be possible to measure the ${}^{131}$Xe precession
frequency in free induction decay in the presence of two other noble gas masers. There are no other
stable noble gas isotopes having non-zero nuclear spin.

Contact shift ratios would be converted to absolute values for enhancement factors by using the
measured values for ${}^{3}$He/Rb \cite{romalis98} and ${}^{3}$He/K \cite{baranga98}. The
enhancement factor for ${}^{3}$He/Cs has not yet been measured.

Measuring contact shift ratios to a few parts in
$10^4$ could proceed as follows: one would first measure the ratio of the enhancement
factors of ${}^{83}$Kr and ${}^{129}$Xe to each other. The contact shift enhancement
factors for ${}^{83}$Kr and
${}^{129}$Xe are estimated by Walker to be about 230 and 730, respectively \cite{walker89}. Using
an independent magnetometer with the colocated Xe-Kr masers, the ratio of the [large] enhancement
factors of Xe and Kr could be measured to a few parts in
$10^4$ even if the classical magnetization field were accounted for only to the $10\%$ level
(which can be done trivially- a $\sim 1\%$ accounting is claimed in this work). Triple masers using
Kr and Xe, along with
${}^{21}$Ne and
${}^{3}$He in turn would yield high-precision measurements of the ratios of all the enhancement
factors; the known enhancement factor for He would then determine the absolute values of the other
enhancement factors. Temperature dependences could of course also be measured. In particular,
enhancement factor ratios as a function of density (normalized to some experimental reference
density) could be determined to very high precision. Improved cell thermometry would eliminate
the need to depend on a temperature-density calibration \cite{alcock84} to determine the cell
temperature. Also, for very high precision measurements, control of the noble gas magnetization
fields would have to be maintained more carefully than in the present work; in particular, the
time between polarization reversals could be decreased in order to increase the relative
size of the contact shifts in comparison to noble gas polarization induced frequency shifts. It
might also be important to account carefully for the contribution of van der Waals molecule
formation to the contact frequency shifts.

Finally, we remark that a two species maser using ${}^3$He and ${}^{21}$Ne has been proposed
for a high precision test of local Lorentz invariance (LLI) \cite{stoner99}, in which Rb-noble
gas contact shifts would be an important systematic effect. Adding a co-located ${}^{129}$Xe
maser would provide a high precision \emph{in situ} measurement of the Rb magnetization. With
part-in-$10^4$ knowledge of the enhancement factors, the systematic effects of contact shifts on
a search for LLI violation could be rendered negligible with the use of the ${}^{129}$Xe maser
"co-polarimeter".

We thank T. Chupp for the use of the ${}^{21}$Ne/${}^{3}$He cell, and gratefully acknowledge D.
Phillips and D. Bear for useful conversations.


\begin{references}
\bibitem{bouchiat60}M.A. Bouchiat, T.R. Carver, and C.M. Varnum,  Phys. Rev. Lett.  {\bf 5},
373 (1960)
\bibitem{walker97}T. Walker and W. Happer, Rev. Mod. Phys.  {\bf 69}, 629 (1997)
\bibitem{abe97}K. Abe \emph{et. al.}, Phys. Rev. Lett. {\bf 79}, 26 (1997); Phys. Lett. B {\bf
404}, 377 (1997)
\bibitem{greene95}G.L. Greene, A.K. Thompson, and M.S. Dewey, Nucl. Inst. Meth. A {\bf 356},
177 (1995)
\bibitem{ng_imaging_i}M.S. Albert, G.D. Cates, B. Driehuys, W. Happer, B. Saam, C.S.
Springer Jr., and A. Wishnia, Nature {\bf 370}, 199 (1994); H. Middleton, L.W.
Hedlund, G.A. Johnson, K. Juvan, J. Schwartz, Magn. Reson. Med. {\bf 33}, 271 (1995)
\bibitem{ng_imaging_ii}G. R. Davies, T. K. Halstead, R. C. Greenhow, and K. J. Packer, Chem.
Phys. Lett. {\bf 230}, 239 (1994); B. R. Patyal et al., J. Magn. Reson. {\bf 126}, 58
(1997); M. Bock, Magn. Reson. Med. {\bf 38}, 890 (1997); D. M. Schmidt et al.,
J. Magn. Reson. {\bf 129}, 184 (1997); H.U. Kauczor, R. Surkau, T. Roberts, Euro. Radiol. {\bf
8}, 820 (1998), and references therein.
\bibitem{ng_imaging_iii}C.H. Tseng, G.P. Wong, V.R. Pomeroy, R.W. Mair, D.P. Hinton, D.
Hoffmann, R.E.  Stoner, F.W. Hersman, D.G. Cory, and R.L. Walsworth, Phys. Rev. Lett. {\bf 81},
3785 (1998)
\bibitem{mair98}R.W. Mair, D.G. Cory, S. Peled, C.H. Tseng, S. Patz, R.L. Walsworth, J. Magn.
Reson. {\bf 35}, 478 (1998); R.W. Mair, G.P. Wong, D. Hoffmann, M.D. Hurlimann, S. Patz, L.M.
Schwartz, R.L. Walsworth, Phys. Rev. Lett. {\bf 83}, 3324 (1999).
\bibitem{chupp94}T.E. Chupp, R.J. Hoare, R.L. Walsworth, and Bo Wu, Phys. Rev. Lett. {\bf 72},
2363 (1994)
\bibitem{stoner96}R.E. Stoner, M.A. Rosenberry, J.T. Wright, T.E. Chupp, E.R. Oteiza,
and R.L. Walsworth, Phys. Rev. Lett. {\bf 77}, 3971 (1996)
\bibitem{romalis99}M.V. Romalis and W. Happer, Phys. Rev. A {\bf 60}, 1385 (1999)
\bibitem{bear98}D. Bear, T.E. Chupp, K. Cooper, S. DeDeo, M. Rosenberry, R.E. Stoner, and R.L.
Walsworth,  Phys. Rev. A {\bf 57}, 5006 (1998)
\bibitem{chupp90}  T.E. Chupp and R.J. Hoare, Phys. Rev. Lett. {\bf 64}, 2261 (1990)
\bibitem{chupp89} T.E. Chupp, R.J. Hoare, R.A. Loveman, E.R. Oteiza, J.M.
Richardson, M.E. Wagshul, A.K. Thompson, Phys. Rev. Lett. {\bf 63}, 1541 (1989)
\bibitem{vold84} T.G. Vold, F. Raab, B. Heckel, and E.N. Fortson, Phys. Rev. Lett. {\bf 52},
2229 (1984)
\bibitem{kostelecky99} V.A. Kostelecky and C.D. Lane, Phys. Rev. D {\bf 60}, 116010 (1999)
\bibitem{stoner99} R.E. Stoner, in \emph{CPT and Lorentz Symmetry}, V.A. Kostelecky, ed.,
World Scientific, Singapore (1999), p. 201
\bibitem{romalis98} M.V. Romalis and G.D. Cates, Phys. Rev. A {\bf 58}, 3004 (1998)
\bibitem{baranga98} A. Ben-Amar Baranga, S. Appelt, M.V. Romalis, C.J. Erickson, A.R. Young, G.D.
Cates, and W. Happer, Phys. Rev. Lett. {\bf 80}, 2801 (1998)
\bibitem{walker89} T.G. Walker, Phys. Rev. A {\bf 40}, 4959 (1989)
\bibitem{happer87} W. Happer and W.A. van Wijngaarden, Hyperfine Int. {\bf 38}, 435 (1987)
\bibitem{happer72} W. Happer, Rev. Mod. Phys. {\bf 44}, 169 (1972)
\bibitem{schaefer89} S.R. Schaefer, G.D. Cates, Ting-ray Chien, D. Gonatas, W. Happer,
and T.G. Walker, Phys. Rev. A {\bf 39}, 5613 (1989)
\bibitem{chupp_pc} T.E. Chupp, private communication
\bibitem{zu88} W. Zu, S.R. Schaefer, G.D. Cates, and W. Happer, Phys. Rev. A {\bf 37}, 1161
(1988)
\bibitem{alcock84} C.B. Alcock, V.P. Itkin, and M.K. Horrigan, Canadian Metallurgical Qtrly {\bf
23}, 309 (1984)
\bibitem{hoffman_kunze} K. Hoffman and R. Kunze, \emph{Linear Algebra}, Prentiss-Hall, Inc.
(1971), p. 280

\end{references}
\end{document}